\documentstyle[multicol,aps,pre,epsf]{revtex}

\begin{document}
\def\f{\phi}
\def\a{\alpha}
\def\b{\beta}
\def\d{\delta}
\def\r{\rho}
\def\l{\lambda}
\def\g{\gamma}
\def\G{\Gamma}
\def\D{\Delta}
\def\m{\mu}
\def\n{\nu}
\def\si{\sigma}
\def\s{|\nabla\r|^2}
\def\rv{\vec r}
\def\kv{\vec k}
\def\p{\psi}
\def\H{{\cal H}}
\def\v{\left.{\d \H[\r] \over\d \r(z)}\right|_{\r=\r_0}\!\!\!\!}
\def\vs{\left.{\d^2 \H[\r] \over\d \r(z_1) \d \r(z_2)}\right|_{\r=\r_0}\!\!\!\!}
\def\vt{\left.{\d^3 \H[\r] \over\d \r(z_1)\d \r(z_2)\d \r(z_3)}\right|_{\r=\r_0}\!\!\!\!}
\def\vf{\left.{\d^4 \H[\r] \over\d \r(z_1)\d \r(z_2)\d \r(z_3)\d \r(z_4)}
\right|_{\r=\r_0}\!\!\!\!}

\title{CONVERGENT APPROXIMATION FOR THE 2-BODY CORRELATION FUNCTION IN AN INTERFACE}

\author{Iaroslav Ispolatov}
\address{
Chemistry Department, Baker Laboratory, Cornell University, Ithaca,
NY 14853, USA}  
 
\date{\today}

\maketitle

\begin{abstract}
A convergent approximation is proposed 
for a mean field density-density
correlation function in a system with a two-phase interface.
It is based on a fourth-order expansion of the Hamiltonian 
in terms of fluctuations around the equilibrium profile.
The approach is illustrated by one and three dimensional calculations
for systems characterized by the Ginzburg-Landau functional.
\end{abstract}

\section{Introduction}
Density correlations in the vicinity of equilibrium planar interfaces
have been extensively studied by numerous authors \cite{e1,e2,w,s,dj,z}.  
Several methods, including the density functional theory
\cite{e1,e2}, the capillary wave model \cite{w,s},
and the eigenstate expansion of the fluctuations \cite {e2,dj,z} were used to calculate the 
correlation function.
Most existing approaches have to confront a problem of divergence, 
caused
by the vanishing energy cost of rigid shifts of the interface in an 
infinite system without external potential.
We will focus in this work on 
the Hamiltonian second derivative $\vs$
eigenstate expansion method (sometimes also called ``field theoretical method'', \cite{e2,z}).
In the framework of this method the divergence manifests itself 
in the term $\f_0(z_1)\f_0(z_2)/\l_0$
where  $\f_0(z)$ and $\l_0$ are the ``zero'' eigenstate and eigenvalue; 
$\f_0(z)\equiv d\r_0(z)/dz$ where $\r_0(z)$ is the equilibrium density profile. 
The eigenvalue $\l_0$ is zero for vanishing external 
localizing field,
The traditional physical explanation for the divergence is the following: 
the density fluctuations corresponding to the interface 
shifts, to the first order of shift amplitude, have component 
only along the zero eigenstate,
$\D\r(z)= \r_0(z+\D z)-\r_0(z)=d\r(z)/dz \D z +{\cal O}(\D z)^2$,
and 
free wandering of the interface as a whole results in the ambiguity
in defining the density-density correlation function. 
In the framework of the capillary wave model (\cite{w}) this ambiguity
is overcome by renormalizing the equilibrium density profile 
by taking into account the free wandering
mode. As a result, average width of the renormalized interface diverges
for vanishing external field, and the zero-order term contribution to the 
density-density correlation function tends to zero.
However, after such renormalization,  one loses information about instantenous 
and local 
fluctuations of the interface.

In an attempt to resolve this ambiguity, 
we would like to focus our attention on the question of divergence of the zero eigenstate term
The main motivation of our approach lies in the following.

It is true that in an infinite system without a confining external field the
energy cost of rigid shifts of the planar two-phase interface is zero,
$\H[\r(z_0+\D z)]=\H[\r(z_0)]$. Yet this does not mean that the amplitude 
$a_0$ of the
density fluctuation proportional  to the zero eigenstate,
$\d \r (z)=a_0 \r'_o(z)$ can grow infinitely without a free-energy 
penalty. Physically it follows from the fact, that although the position of the interface is undetermined, 
the actual values of the density near the interface 
cannot go significantly beyond the density of either of the bulk phases.
For all the realistic Hamiltonian functionals describing stable
systems, this non-divergence is controlled by 
a positive coefficient in front of the highest power of the density.
For the Ginzburg-Landau 
functional 
\begin{equation}
\label{GL}
\H[\r]=\int (\{{d\over dz}\r(z)\}^2+\{1-\r^2(z)\}^2) dz
\end{equation}
the latter is $+\r^4(z)$. The divergence that appears when only the 
harmonic (second-order) terms of the  
expansion of the Hamiltonian are used is the direct consequence of
neglect of all higher-order terms.
Hence, to eliminate the divergence of the zero-eigenstate term in the correlation 
function, 
it is natural to try using
higher-order terms of the expansion of the Hamiltonian around the 
equilibrium density profile (see \cite{j} where the equilibrium profile was calculated using 
higher-order terms).
It turns out that fourth-order terms are sufficient to keep
the zero-eigenstate contribution finite.
We propose an approximate method to consider these fourth order terms;
this method is formally introduced in Section II. Concrete examples for 
the Ginzburg-Landau Hamiltonian for one- and three- dimensional systems 
are presented in Sections III and IV. 

However, after eliminating the divergence of the zero-order eigenstate term, it
is natural to ask,
which eigenstate or combination of eigenstates of the second 
derivative matrix describe the macroscopic shifts of the interface 
$\D\r(z)= \r_0(z+\D z)-\r_0(z)$?
The expansion coefficient of this density fluctuation along the zero 
eigenstate remains finite
even for infinite shifts,
\begin{equation}
\label{gen1}
\int_{-\infty}^{+\infty}\D\r(z)\r_0'(z)dz \rightarrow \r_0(+\infty)
[\r_0(+\infty)-\r_0(-\infty)]
\end{equation}
for $\D z\rightarrow +\infty$.
This statement could serve as another argument for a finite average value of the 
amplitude of fluctuation along 
$\r_0'(z)$. The same is true for all other bound (localized) eigenstates.
Yet the projection of the  shift $\r_0(z+\D z)-\r_0(z)$ onto low-lying 
continuum states
diverges as $\D z \rightarrow \infty$. For the Ginzburg-Landau Hamiltonian 
the mean-field
equilibrium density profile is $\r_0(z)=\tanh(z)$, and the 
first non-localized eigenstate of the second derivative matrix is proportional to
$(3  \tanh(z)-1)/2$. The integral
\begin{equation}
\label{gen}
\int_{-\infty}^{+\infty}(\tanh(z+\D z)-\tanh(z))(3/2  \tanh(z)-1/2)dz \rightarrow 2 \D z 
\end{equation}
diverges linearly for $\D z \rightarrow \infty$. 
It means that since macroscopic shifts have no energy cost, the appropriate 
linear combination of low-lying continuum states with some of the 
coefficients growing proportionally to the 
magnitude of the shift also has no energy cost. The finiteness 
of the average values of all the continuum spectrum  amplitudes 
is another artifact of second-order truncation; in particular,
it is a result of the neglect of the mixing of different harmonics in the third 
and 
higher-order terms.
Consequently, our approximation is convergent only because, after going to the 
higher-order expansion in the zero term, we stopped short of going beyond the harmonic
approximation for terms containing the lowest-lying continuum states.
 
However, the main contribution to the correlations near the interface comes from the 
bound states, while continuum states are more relevant for the correlations in the 
bulk phases.
A possible merit of our approximation is the improvement in the accuracy
of 
density correlation calculations
in the vicinity of the interstate. To compare our results to experimental data, theoretical calculations should be supplemented by {\it a priori} knowledge of 
the macroscopic localization of the interface.

For a three-dimensional system the situation is similar. For 
square-gradient energy functionals, the perturbation
$\D\r(x,y,z)=\r_0(z+f(x,y))-\r_0(z)$ of an initially flat interface
has the energy cost 
\begin{equation}
\label{3di}
\D F = \int_{-\infty}^{+\infty}
[\r'_0(z)]^2 dz\int_{-\infty}^{+\infty}\int_{-\infty}^{+\infty} |\nabla f(x,y)|^2
dx dy
\end{equation}
For long-wavelength fluctuations, $f(x,y)\sim\exp[i(k_x x+k_y y)]$,
$|\vec k|\ll 1$, the energy cost of such fluctuations vanishes as
$|k|^2$. However, considering the expansion of the density fluctuations over
the system of eigenstates (now in 3D) of the Hamiltonian second derivative matrix,
$\psi_i(\vec r)\equiv 1/L\; \f_i(z) \exp(ik_x x)\exp(ik_y y)$, the divergent 
contribution comes not from the terms with localized $\f_i$, but from the bottom 
of the continuum
of $z$-coordinate eigenstates. To illustrate that, in section IV we 
calculate the convergent 
contribution to the correlation function from the $\f_0(z) \exp(ik_x x)\exp(ik_y y)$ 
eigenstate.

\section{Formalism}
A density-density correlation function $g(z_1,z_2)$ 
is defined as a thermal average of a product of density fluctuations 
$\D \r(z)=\r(z)-\r_0(z)$
around the equilibrium density 
profile $\r_0(z)$:
\begin{equation}
\label{cor}
g(z_1,z_2)\equiv \langle\D\r(z_1)\D\r(z_2)\rangle.
\end{equation}
For simplicity, in this section we consider a one-dimensional case, $\r=\r(z)$.
Following \cite{z} we express it as a functional integral over all possible density
profiles,
\begin{equation}
\label{def}
 g(z_1,z_2)=(1/Z)\int { \cal D}\r \; \D\r(z_1)\D\r(z_2) \exp\{ -{\H[\r]}\}
\end{equation}
where ${\H[\r]}\equiv H[\r]/k_b T$ is a reduced Hamiltonian functional,
and the partition function $Z$ serves as the normalization constant,
\begin{equation}
\label{nor}
 Z=\int { \cal D}\r \; \exp\{ -{\H[\r]}\}
\end{equation}

In the mean-field approximation that we will use through this work, 
the equilibrium density profile
$\r_0(z)$ 
is determined as the one that
minimizes the Hamiltonian,
\begin{equation}
\label{eqi}
\v=0.  
\end{equation}
To evaluate (\ref{def}),
we proceed by calculating the eigenstates of the integral operator
with the kernel $\vs$,
\begin{equation}
\int \vs \f_i(z_2)\,dz_2=\l_i\f_i(z_1).  
\label{eig}
\end{equation}
There is always a special eigenstate corresponding to $\f_0(z)=d\r_0(z)/dz$
which has zero eigenvalue $\l_0=0$ since
\begin{equation}
\label{zer}
\int \vs {d\r_0(z_2)\over dz_2}\, dz_2={d\over dz}\v \equiv0. 
\end{equation}  

The Hamiltonian is usually real and contains only even powers
of differential operators; it makes the integral operator in Eq.~(\ref{eig})
Hermitian. The
system of eigenstates $\f_i$ is complete 
and orthogonal; we also assume that it is normalized with unit weight function
and all eigenfunctions $\f_i$ are made real.
An arbitrary density fluctuation $\D\r(z)= \r(z)-\r_0(z)$
can be expanded over the complete  set of functions $\{\f_i(z)\}$:
\begin{equation}
\label{exp}
\D\r(z)=\sum_{i=0}^{\infty} a_i \f_i(z), \; \; a_i=\int \D\r(z)\f_i(z)dz
\end{equation}
Using the expansion(\ref{exp}), the functional integral (\ref{def}) can be expressed
 as
\begin{equation}
\label{cor2}
g(z_1,z_2)={1\over Z}\sum_{i=0}^{\infty} \sum_{j=0}^{\infty} \int\ldots \int
a_i a_j \f_i(z_1) \f_j(z_2) \exp \{-\H[\r_0+\sum_{k=0}^{\infty}a_k \f_k]\}
\prod_{m=0}^{\infty}da_m
\end{equation} 
where the normalization constant 
\begin{equation}
\label{cor3}
Z=\int\ldots \int
\exp \{-\H[\r_0+\sum_{k=0}^{\infty}a_k \f_k]\}
\prod_{m=0}^{\infty}da_m.
\end{equation} 
We assume here that the integrals in both numerator and denominator 
are convergent. As we mentioned in the Introduction, it is not true
for the single specific direction in the space of coefficients $\{a_i\}$, 
corresponding to
the rigid macroscopic shifts of the interface. However, as a result of the 
approximations made below, this divergence will not affect the further 
calculations.


Traditionally, $\H[\r_0+\sum_{k=0}^{\infty}a_k \f_k]$ 
is expanded to the second order around the 
equilibrium density profile, the orthogonality conditions for the $\f_i$ are 
used, 
and the corresponding Gaussian integrals factorized \cite{e2,z}. 
As a result, the familiar expression for the 
density-density correlation function is 
recovered:
\begin{equation}
\label{trad}
g(z_1,z_2)=\sum_{i=0}^{\infty}\langle a_i^2\rangle\f_i(z_1) \f_i(z_2),
\end{equation} 
where $\langle a_i^2\rangle={1/ \l_i}$.
As we already mentioned, the zero term   
diverges since $\l_0=0$.

However, as we discussed in the Introduction, from physical considerations 
$\langle a_0^2\rangle$ must 
have finite value. It is indeed the case
if in Eqs.~(\ref{cor2}), (\ref{cor3}) one goes to higher than second order in 
the expansion of $\H[\r]$.  

In our case, the expansion of  
$\H[\r]$ up to the fourth order in the density fluctuation around the 
equilibrium profile is sufficient:
\begin{eqnarray}
\label{big}
\nonumber
&\H[\r_0+{\displaystyle\sum_{k=0}^{\infty}}a_k \f_k]\approx &\\
\nonumber
&{\displaystyle\sum_{k=0}^{\infty}a_k \int }\v \f_k(z) 
dz+&\\
&{1\over2}{\displaystyle \sum_{k,l=0}^{\infty}a_k a_l\int\int}\vs\f_k(z_1)
\f_l(z_2) {\displaystyle \prod_{j=1}^2 }dz_j +&\\
\nonumber
&{1\over3!}{\displaystyle \sum_{k,l,m=0}^{\infty}a_k a_l a_m
\int\int\int} \vt \f_k(z_1)\f_l(z_2)\f_m(z_3){\displaystyle \prod_{j=1}^3}
 dz_j  +&\\
\nonumber
&{1\over4!}{\displaystyle \sum_{k,l,m,n=0}^{\infty}a_ka_l a_m a_n 
\int\int\int\int} \vf
\f_k(z_1)\f_l(z_2)\f_m(z_3)\f_n(z_4) {\displaystyle \prod_{j=1}^4 dz_j}.& 
\end {eqnarray}
The first order term is identically zero, the second-order terms are used in the 
traditional
formalism, and the third and fourth-order terms are essential for our treatment. 
If we substitute this expansion into the expressions for the correlations 
function (\ref{cor2}), (\ref{cor3}), for non-pathological forms of the 
Hamiltonian 
functional $\H[\r]$, it will produce a finite value for  
$\langle a_0^2\rangle$. However, in their complete form Eqs.~(\ref{cor2}),
 (\ref{cor3}), (\ref{big}) are
hardly tractable. Assuming that we are far enough from a critical point, we
use the following approximation to evaluate 
$\langle a_0^2\rangle$.
We assume that the second-order expansion works well enough 
for all $\langle a_i^2\rangle={1/ \l_i}$ with $i\neq 0$ 
and drop from Eq.~(\ref{big}) 
all terms
that do not contain $a_0$. In the remaining terms we replace
all combinations of $a_i$, $a_ia_j$, $a_ia_ja_k$, 
$\{i,j,k\}\neq0$ by their average values obtained with
the second-order expansion:
\begin{eqnarray}
\label{av}
\nonumber
&\langle a_i \rangle=0,&\\
&\langle a_i a_j \rangle=\d_{ij}{1\over \l_i},&\\
\nonumber
&\langle a_i a_j a_k \rangle=0.&\\
\nonumber
\end{eqnarray} 
This approximations allow us to express $\langle a_0^2\rangle$ in the form of the 
following integral:
\begin{equation}
\label{e}
\langle a_0^2\rangle={\int x^2 \exp\{-\alpha x^4 -\beta x^2 -\gamma x -\d x^3
\}dx \over
\int \exp\{-\a x^4 -\b x^2 -\g x -\d x^3\} dx},
\end {equation}
with coefficients $\a,\;\b,\;\g$ given by
\begin{eqnarray}
\label{coeff}
\nonumber
&\alpha={1\over4!}{\displaystyle 
\int\int\int\int} \vf
\f_0(z_1)\f_0(z_2)\f_0(z_3)\f_0(z_4){\displaystyle  \prod_{j=1}^4 }dz_j,&\\
&\beta={1\over4}{\displaystyle \sum_{i=1}^{\infty}
{1\over\l_i}\int\int\int\int \vf
\f_i(z_1)\f_i(z_2)\f_0(z_3)\f_0(z_4) \prod_{j=1}^4 }dz_j,&\\
\nonumber
&\gamma={1\over2}{\displaystyle \sum_{i=1}^{\infty}{1\over\l_i}
\int\int\int \vt
\f_i(z_1)\f_i(z_2)\f_0(z_3) \prod_{j=1}^3 }dz_j,&\\
\nonumber
&\d={1\over3!}{\displaystyle 
\int\int\int \vt
\f_0(z_1)\f_0(z_2)\f_0(z_3) \prod_{j=1}^3 }dz_j.&\\
\nonumber
\end{eqnarray}

\section{Ginzburg-Landau Hamiltonian}

To illustrate our approach let us a consider a simple
1D system with the Ginzburg-Landau Hamiltonian (\ref{GL}). The
equilibrium density profile for symmetric boundary conditions
$\r(-l)=-1$, $\r(+l)=1$, $l\rightarrow \infty$ is
$\r_0(z)=\tanh(z)$. The eigenvalue equation (\ref{eig}) takes the 
differential form
\begin{equation}
\label{shr}
-2{d^2\over dz^2}\f_i(z)-4\f_i(z)+12\tanh^2(z)\f_i(z)=\l_i\f_i(z),
\end{equation}
and the third and fourth-order functional derivatives of $\H[\r]$ are:
\begin{equation}
\vt=24\tanh(z_1)\d(z_1-z_2)\d(z_2-z_3)
\end{equation}
\begin{equation}
\vf=24\d(z_1-z_2)\d(z_2-z_3)\d(z_3-z_4).
\end{equation}
In fact, the expansion (\ref{big}) is now exact for $\H[\r]$ 
since
all higher-order variational derivatives are identically equal to zero.
To proceed further we need to calculate 
$\a$, $\b$, $\g$, and $\d$ as defined in
Eq.~(\ref{coeff}). From parity consideration, 
the coefficients $\g$ and 
$\d$ are zero.

The calculation of $\a $ is straightforward:
\begin{equation}
\label{a}
\a= {9\over 16} 
\int_{-\infty}^{+\infty}{dz \over \cosh^{8}(z)}
={18\over35}.
\end{equation}
The coefficient ${9\over 16}$ appears from the normalization condition:
\begin{equation}
\label{norm}
\int_{-\infty}^{+\infty}\f_i^2(z)dz=1.
\end{equation}
To evaluate $\b$ we need to know the eigenstates $\f_i(z)$ of
Eq.~(\ref{shr}):
\begin{equation}
\label{b}
\b={9\over 2}\sum_{i=1}^{\infty}\int_{-\infty}^{+\infty}
{\f_i^2(z)\over\l_i}
{dz \over \cosh^{4}(z)}.
\end{equation}
The sum in Eq.~(\ref{b}) contains a contribution from one bound state,
($\f_1(z)=\sqrt{3/2} \sinh(z)\cosh^{-2}(z)$, $\l_1=6$) and continuum states 
($i\geq 2$)
To evaluate the sum over the continuum, 
we use expressions obtained in \cite{z} (Eq.~(3.15 - 3.20)):
\begin{equation}
\label{cont}
{9\over 2}\sum_{i=2}^{\infty}\int_{-\infty}^{+\infty}
{\f_i^2(z)\over\l_i}
{dz \over \cosh^{4}(z)}={9\over 4}\int_{-\infty}^{+\infty}
{dk \over 2 \pi}{1 \over (k^2+4)^2}{1 \over (k^2+1)}
\int_{-\infty}^{+\infty}{\f_k(z) \over \cosh^{4}(z)} dz  
\end{equation}
where the continuum eigenstates $\f_k(z)$ are
\begin{equation}
\label{fk}
\f_k(z)=\exp(ikz)[1+k^2+3ik\tanh(z)-3\tanh^2(z)].
\end{equation}
The normalization of $\f_k(z)$ is taken into account in the $k$-dependent factor
of the outer integral in (\ref{cont}).
Both integrations in (\ref{cont}) are straightforward and finally we obtain for
$\b$

\begin{equation}
\label{be} 
\b\approx {6\over35} + 0.022 \approx 0.193 
\end {equation}
The contribution from the continuum to the value of $\b$ is $\approx 13\%$. 
We substitute these values of $\a$ and $\b$ into Eq.~(\ref{e}), and, in the first order
in $\b/ \sqrt{\a}$, obtain for 
$\langle a_0^2\rangle$: 
\begin{equation}
\label{et}
\langle a_0^2\rangle \approx {\int x^2 \exp\{-\a z^4\}(1-\b z^2) dz \over
\int \exp\{- {\a}z^4\} (1-\b z^2) dz}={1\over \sqrt{\a}} \; {\G^2({3\over 4})-
{\b \over \sqrt{\a}}{\pi \sqrt{2}\over 4} \over \pi \sqrt{2} -{\b \over \sqrt{\a}}
\G^2({3\over 4})} \approx 0.415
\end {equation} 
It is interesting to note that if one ignores all cross-terms (
contribution from $\b$) in
Eq.~(\ref{e}), $\langle a_0^2\rangle \approx 0.471$.
For the Ginzburg-Landau Hamiltonian, ${\b \over \sqrt{\a}}\approx
0.27$ plays the role of a small parameter in
the approximations (\ref{av}) - (\ref{coeff}), 
as well as in (\ref{et}).
Hence there is a certain degree of numerical justification of heuristic assumptions made in (\ref{av}) - (\ref{coeff}).
To obtain the final result for the density-density correlation function 
(\ref{trad})
we use the following method. When the external potential $V(z)=cz$, 
linear in the coordinate $z$ measured form the interface location
$z=0$, 
is added to the Hamiltonian, the zero eigenvalue
becomes proportional to the coefficient in front of this term:
$\l_0 \propto c$.
In \cite{z} an expression for the correlation function 
$G_c(z_1,z_2)$  (Eq. (4.17) in \cite{z})
is obtained in the presence of an external potential $V(z)$, with 
$V(z)\rightarrow 
c\tanh(z)$ when $c \rightarrow 0$.
It is shown that the zero eigenvalue $\l_0/c \rightarrow 1 $ for 
$c \rightarrow 0$. Since for $i\geq 1$ all  $\l_i$ go to constant limits when
the external potential is turned off, the ``truncated'' correlation function 
$\bar g$
(without the zero eigenstate term) can be expressed as:
\begin{equation}
\label{gbar}
\bar g({z_1,z_2})\equiv\sum_{i=1}^{\infty}{\f_i(z_1) \f_i(z_2)\over  \l_i}=
\lim_{c \rightarrow 0} {d \over d c}cG_c(z_1,z_2).
\end {equation}
To recover the ``non-truncated'' correlation function $g({z_1,z_2})$ we
add to $\bar g({z_1,z_2})$  from (\ref{gbar}) the correct contribution form the
zero-order term,
\begin{equation}
\label{gr}
g({z_1,z_2})= \bar g({z_1,z_2})+{3\over 4\cosh^2(z_1)\cosh^2(z_2)}
\langle a_0^2\rangle. 
\end {equation}
 with $\langle a_0^2\rangle$ given by (\ref{et}). 
Sketches of the density-density correlation function (\ref{gr})
are presented in Figs.~1,2.
It is straightforward to demonstrate that far from the interface 
(${z_1.z_2}\gg 1$), the density-density correlation function
(\ref{gr}) goes to the correct bulk phase limit,
\begin{equation}
\label{bulk}
g_{bulk}(z_1,z_2)={\exp(-2|z_1-z_2|)\over 8}.
\end{equation}

\begin{figure}
\centerline{\epsfxsize=12cm \epsfbox{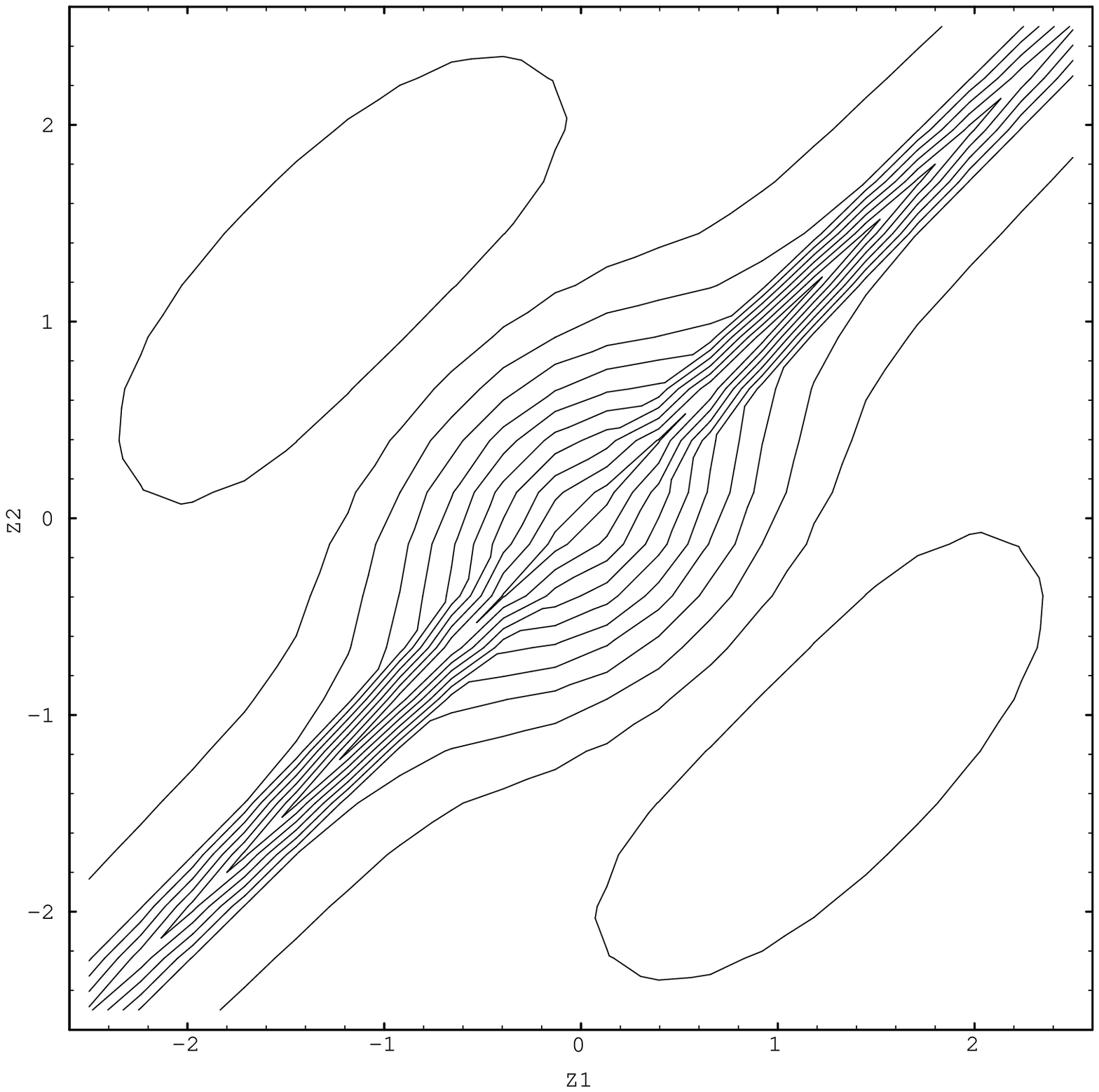}}
\centerline{{\small {\bf Fig.~1}. 
A contour plot of the density-density correlation function
$g(z_1,z_2)$.}}
\end{figure}
\begin{figure}
\centerline{\epsfxsize=12cm \epsfbox{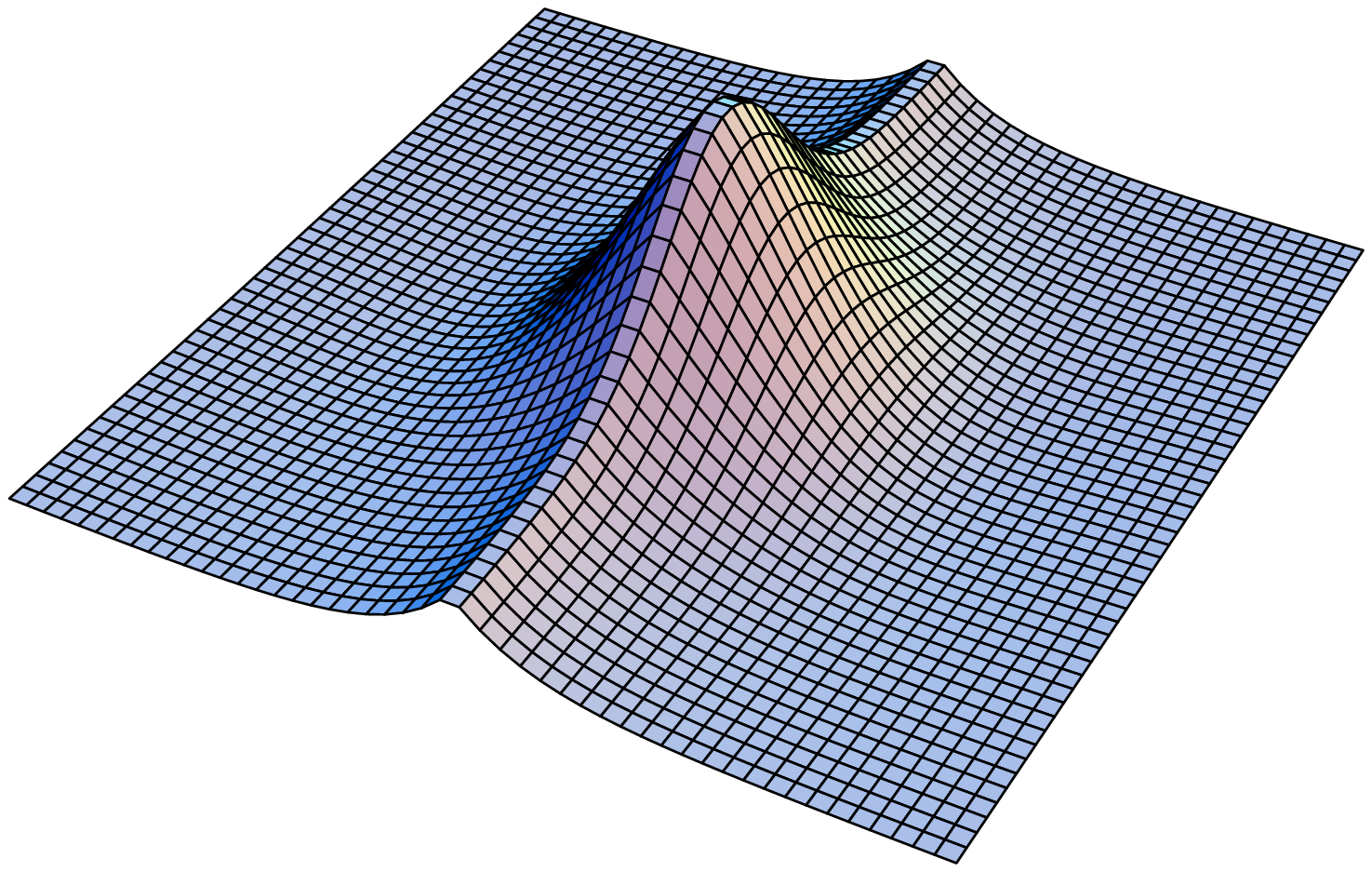}}
\centerline{{\small {\bf Fig.~2}. 
A 3D sketch of the density-density correlation function
$g(z_1,z_2)$.}}
\end{figure}
\section{3D Calculation}
The next logical step is to generalize this approach to
 a more realistic 3D system.
For simplicity, we consider the same 
Ginzburg-Landau Hamiltonian as in (\ref{GL})
\begin{equation}
\label{GL3D}
\H[\r]=\int (\s+\{1-\r^2(\rv)\}^2 )d\rv
\end{equation}
The equilibrium density profile for the symmetric boundary conditions
$\r(z=-l)=-1$, $\r(z=+l)=1$, $l\rightarrow \infty$ is the same 
as in the 1D case,
$\r_0(z)=\tanh(z)$. The
three-dimensional analog of the 
eigenvalue equation (\ref{shr}) reads
\begin{equation}
\label{shr3D}
-2\D\p_i(\rv)-4\p_i(\rv)+12\tanh^2(z)\p_i(\rv)=\l_i\p_i(\rv),
\end{equation}
Since the ``potential'' term in (\ref{shr3D}) depends only on $z$,
the variables can be separated, $\p_i(\rv)=\f_i(z)\xi_{i}(x,y)$.
We assume that in the $xy$ plane the system is confined to a square box of
size
$L$ with periodic boundary conditions. 
Then the  $xy$ components of the eigenstates of (\ref{shr3D}) 
are the normalized plane waves 
$\xi_{\kv}(x,y)=(1/L)\exp(ik_x x)
\exp(ik_y y)$ with $ k_{\{x,y\}}=2\pi n_{\{x,y\}}/L$, $ n_{\{x,y\}}=0,\pm 1,
\pm 2 \ldots$. 
Expanding density variations over the complete set of
functions $\p_{i\kv}(\rv)$ and taking the thermal average, 
we obtain for the density-density correlation function (compare to (\ref{cor2})):
\begin{eqnarray}
\label{cor3D}
\nonumber
&g(\rv_1,\rv_2)={1\over Z}\displaystyle\sum_{i,\kv_1} 
\displaystyle\sum_{j,\kv_2} \int \ldots \int
a_{i.\kv_1}
 a_{j\kv_2} \p_{i\kv_1}^*(\rv_1) \p_{j\kv_2}(\rv_2)&\\
& \exp \{-\H[\r_0+\displaystyle\sum_{m,\kv_3} a_{m\kv_3} \p_{m\kv_3}]\}
\displaystyle \prod_{n,\kv_4}da_{n\kv_4}.&
\end{eqnarray}
Similarly to (\ref{cor3}),  $Z$ is a normalization constant:
\begin{equation}
\label{cor33D}
Z=\int \ldots \int
\exp \{-\H[\r_0+\sum_{m,\kv_1} a_{m\kv_1} \p_{m\kv_1}]\}
\prod_{n,\kv_2}da_{n\kv_2}.
\end{equation} 
Here and below we use the shorthand notation: for sums, products, and 
subscript indexes symbol $\kv$ means the set of $xy$ eigenstate
indexes $\{n_x,n_y\}$.
After expanding (\ref{cor3D}) to the second order in the 
density variation,
an expression analogous to  (\ref{trad}) is recovered:
\begin{equation}
\label{trad3D}
g(\rv_1,\rv_2)=\sum_{i,\kv}\langle a_{i\kv}^2\rangle\p_{i\kv}^*(\rv_1) 
\p_{i\kv}(\rv_2),
\end{equation}
with $\langle a_{i\kv}^2\rangle={1/ (\l_i+2k^2)}$.
As in the (\ref{trad}), a similar problem arises for the 
$\l_0=0$ eigenstate: 
for $L\rightarrow \infty$ the sum on $\kv$ diverges 
at the lower limit.

To avoid the divergence occurring in the $\l_0=0$ term, we suggest the same 
recipe as in the 
one-dimensional 
case: to continue the expansion of the Hamiltonian to 
the fourth order.
For simplicity, we neglect mixing of the eigenstates with different $\l_i$
in the fourth-order term. Similarly to the 1D case, where the relative 
contribution of mixing is given by the ratio of
coefficients $\b/\sqrt{\a}\approx 0.27 $ (\ref{a},\ref{be},\ref{et}),
the inclusion of this 
mixing here will not affect convergence
but will slightly change the numerical values of the coefficients.
For all $\l_i \neq 0$, the second-order result (\ref{trad3D}) is sufficient,
hence we perform fourth-order expansion only for the subspace of 
eigenstates with $\l_0=0$.
The contribution from the $\l_0=0$ eigenstate to the density-density correlation function can be expressed as:
\begin{eqnarray}
\label{cor3D0}
\nonumber
&g_0(\rv_1,\rv_2)={1\over Z_0}\displaystyle\sum_{\kv_1} 
\displaystyle\sum_{\kv_2} \int \ldots \int
a_{0\kv_1}
 a_{0\kv_2} \p_{0\kv_1}^*(\rv_1) \p_{0\kv_2}^*(\rv_2)&\\
& \exp \{-\H[\r_0+\displaystyle\sum_{\kv_3} a_{0\kv_3} \p_{0\kv_3}]\}
\displaystyle \prod_{\kv_4}da_{0\kv_4},&\\
\nonumber
&Z_0=
\displaystyle \int \ldots \int
 \exp \{-\H[\r_0+\displaystyle \sum_{\kv_1} a_{0\kv_1} \p_{0\kv_1}]\}
\;\displaystyle \prod_{\kv_2}da_{0\kv_2}.&
\end{eqnarray}
Using the orthogonality conditions for $\xi_{\kv}(z,y)$ and 
Eq.~(\ref{coeff}) we obtain
\begin{eqnarray}
\label{cor3D1}
\nonumber
&g_0(\rv_1,\rv_2)={1\over Z_0}\displaystyle\sum_{\kv_1} 
 \int \ldots \int
a_{0\kv_1}^2
 \p_{0\kv_1}^*(\rv_1) \p_{0\kv_1}(\rv_2)&\\
& \exp \{-2\displaystyle\sum_{\kv_2} a_{0\kv_2}^2 k_2^2 
-{\a\over L^2}[\displaystyle\sum_{\kv_3} a_{0\kv_3}^2]^2 \}
\;\displaystyle \prod_{\kv_4}da_{0\kv_4},&\\
\nonumber
&Z_0=\displaystyle
 \int \ldots \int
\exp \{-2\displaystyle \sum_{\kv_1} a_{0\kv_1}^2 k_1^2 -
{\a\over L^2}[\displaystyle \sum_{\kv_2}
a_{0\kv_3}^2]^2 \}
\;\displaystyle \prod_{\kv_3}da_{0\kv_3},&
\end{eqnarray}
where $\a$ is defined by (\ref{a}).
Direct evaluation of the functional integrals in (\ref{cor3D1}) is impossible; 
however, a simple approximation will allow us to obtain a physically reasonable
expression for $g_0(\rv_1,\rv_2)$.
The main contribution to the integral in (\ref{cor3D1}) comes from
$a_{0\kv}$ with small $|\kv|$, so in the first approximation 
it is natural to introduce 
an upper cutoff $C$ in sums on $\kv$.
In particular, we replace the infinite limits in all 
the sums and products on $n_x$ and $n_y$ in (\ref{cor3D1}) by 
the finite cutoff,
$|n_{x,y}|\leq \sqrt{C}L/2$. It corresponds to a system-size-independent
cutoff for a wavevector $\kv$ with $|k_{x,y}|\leq \sqrt{C} \pi$.
We select $C$ in such a way that 
the term $-2\sum_{\kv} a_{0\kv}^2 k^2$, quadratic in $a_{0\kv}$,
can be neglected in the exponent, which decouples the integration over
$da_{0\kv}$ from the summation on ${n_x,n_y}$.
Besides neglecting the quadratic term, we remove
the functions $\p_{0\kv}(\rv)$ from (\ref{cor3D1}),
and call the remaining expression  $A$.
\begin{equation}
\label{A}
A \equiv {\displaystyle \sum_{k_1}^N 
 \int \ldots \int
a_{k_1}^2
\exp \{-{\a\over L^2}[\displaystyle\sum_{k_2=1}^N a_{k_2}^2]^2 \}
\;\displaystyle \prod_{k_3}^N da_{k_3}\over
\displaystyle 
 \int \ldots \int
\exp \{-{\a\over L^2}[\displaystyle\sum_{k_4=1}^N a_{k_4}^2]^2 \}
\;\displaystyle \prod_{k_5}^N da_{k_5}},
\end{equation}
with $N=CL^2$.  This expression can be evaluated by using
$N$-dimensional spherical coordinates, $\sum_{k}^N a_{k}^2 \equiv R^2$  
\begin{equation}
\label{A1}
A = {\displaystyle \int R^2 
\exp \{-{\a\over L^2}R^4\}
R^{N-1}\Omega _N dR \over
\displaystyle \int 
\exp \{-{\a\over L^2}R^4\}
R^{N-1}\Omega _N dR }={L\over\sqrt{\a}}{\G({N+2\over 4})\over
\G({N\over 4})}={L\over 2}[\sqrt{N\over\a}+{\cal O}
({1\over\sqrt{N}})]
\end{equation}
Here $\Omega_N=2\pi^{N/2}/\G(N/2)$ is the area of the N-dimensional 
sphere.
By inspection, one can identify $A$ as the sum of the first N terms of
$\langle a^2\rangle $, $A=\sum_{l=1}^N\langle a^2\rangle $.  Therefore
 
\begin{equation}
\label{A2}
\langle a^2\rangle={A\over N}={1\over 2} \sqrt{1\over \a C}
\end{equation}
Now we return to Eq.~(\ref{cor3D1}) and replace one sum
in the fourth-order term in the exponential by 
$\sum_{i=1}^N \langle a^2\rangle$:
\begin{equation}
{\a\over L^2}[\displaystyle\sum_{\kv} a_{0\kv}^2]^2\approx
{\a\over L^2}[\displaystyle\sum_{\kv} a_{0\kv}^2]N \langle a^2 \rangle
={\sqrt{\a C}\over 2}[\displaystyle\sum_{\kv} a_{0\kv}^2]
\end{equation}
After this substitution, Eq.~(\ref{cor3D1}) becomes a product of
Gaussian integrals and its evaluation becomes trivial: 
\begin{equation}
\label{g0}
g_0(\rv_1,\rv_2)\approx \f_0(z_1) \f_0(z_2)
\sum_{\kv}\tilde a_{\kv}^2 {\exp [i (k_x x+ k_y y)]\over L^2},
\end {equation}
where the ``improved'' average values $\tilde a_{\kv}^2$ are given by
\begin{equation}
\label{g1}
\tilde a_{\kv}^2\equiv{ \int a_{\kv}^2 \exp[-a_{\kv}^2(2k^2+\sqrt{\a C}/2)]
d a_{\kv}\over \int \exp[-a_{\kv}^2(2k^2+\sqrt{\a C}/2)]
d a_{\kv}}={1\over 4 k^2+\sqrt{\a C}}   
\end {equation}
Taking the limit $L\rightarrow\infty$ and replacing 
the summation $\sum_{n_x,n_y}$ in 
(\ref{g0}) by the integration
$(L/2\pi)^2\int \int dk_x dk_y$, we obtain
\begin{equation}
\label{gf0}
g_0(\rv_1,\rv_2)\approx \f_0(z_1) \f_0(z_2)
{1\over (4\pi)^2} \int_0^{\infty}{k dk\over k^2 + \sqrt{\a C}/4}
J_0(kr')=\f_0(z_1) \f_0(z_2){1\over (4\pi)^2} K_0[{
(\a C)^{1/4} r_{\perp} \over 2}]. 
\end {equation}
Here $J_0$ and $K_0$ are Bessel and Modified Hankel
functions of zero order and
$r_{\perp}\equiv\sqrt{(x_1-x_2)^2+(y_1-y_2)^2}$.
For large positive $q$ one has $K_0(q)=\sqrt{\pi/2q}\exp(-q)
[1+{\cal O} (1/q)]$. Consequently, we can identify $(\a C)^{1/4}/2$ with the 
previously introduced upper cutoff for the wavevector $k$, {\it i.e.}, with 
$\sqrt{C} \pi$.
It allows us to express the constant $C$ through the known parameters of 
the system,
$C=\a/(4\pi)^4$.
Finally, we write for the zero-eigenstate contribution to the
correlation function:
\begin{equation}
\label{gf}
g_0(\rv_1,\rv_2)\approx \f_0(z_1) \f_0(z_2){1\over (4\pi)^2}
K_0[{\sqrt{ \a}\over 4\pi}\sqrt{(x_1-x_2)^2+(y_1-y_2)^2}]. 
\end {equation}
The contribution from other eigenstates, $\tilde g(\rv_1,\rv_2)$,
is obtained by straightforward integration,
\begin{equation}
\label{go}
\tilde g(\rv_1,\rv_2)= \sum_{j=1}\f_j(z_1) \f_j(z_2){1\over 2 (2\pi)^2}
K_0[\sqrt{ \l_j\over 2}\sqrt{(x_1-x_2)^2+(y_1-y_2)^2}] 
\end {equation}.

\section{Conclusion}

 We show that taking into account the fourth-order terms in the expansion of the 
Hamiltonian
in the calculation of the density-density correlation function makes
the previously divergent (for $D\leq3$) zero-eigenstate term convergent.
We also note that the macroscopic shifts
of the interfacial profile are described not by the zero and other bound 
eigenstates,
but by the combination of low-lying
continuum eigenstates of the Hamiltonian second-derivative matrix.
The inclusion of the convergent zero-order term allows us to 
improve the accuracy of the calculation of the correlation function  
in the vicinity of the interface.
Our approach could be relevant for the experimental results
obtained in the microgravity conditions using a local analytical method, 
{\it e.g.}, scattering with a narrow beam, focused on the interface. 

\section{Acknowledgments}
This work was done in the research group of Prof. B. Widom as part of a program supported by the U.S. National Science Foundation and the  Cornell Center for
Material Research. I thank Prof. Widom for having suggested this problem, 
for stimulating discussions during the course of the work, and for comments 
on the manuscript.




\end{document}